\documentclass[10pt,aip,twocolumn]{revtex4-1}

\usepackage{graphicx}  % needed for figures
\usepackage{dcolumn}   % needed for some tables
\usepackage{bm}        % for math 
\usepackage{amssymb}  
\usepackage{amsmath}
\usepackage{hyperref,color,braket}
\usepackage{epsfig,wrapfig}
\usepackage{gensymb}

\newcommand{\beq}{\begin{equation}}
\newcommand{\eeq}{\end{equation}}

\newcommand{\fr}{\bm{r}}

\begin{document}
\title{High-Quality Resonances in Quasi-Periodic Clusters of Scatterers for Flexural Waves}
\author{Marc Mart\'i-Sabat\'e}
\affiliation{GROC, UJI, Institut de Noves Tecnologies de la Imatge (INIT), Universitat Jaume I, 12071, Castell\'o, Spain}
\author{S\'ebastien Guenneau}
\affiliation{UMI 2004 Abraham de Moivre-CNRS, Imperial College London, London SW7 2AZ, UK}
\author{Dani Torrent}
\email{dtorrent@uji.es}
\affiliation{GROC, UJI, Institut de Noves Tecnologies de la Imatge (INIT), Universitat Jaume I, 12071, Castell\'o, Spain}
\date{\today}
%%%%%%%%%%%%%%%%%%%%%%%%%%%%%%%%%%%%%%%%%%%%%%%%%%%%%%%%%%%%%%%%%%%%%%%%%%%%%%%%%%%%%%%%%%%

%%%%%%%%%%%%%%%%%%%%%%%%%%%%%%%%%%%%%%%%%%%%%%%%%%%%%%%%%%%%%%%%%%%%%%%%%%%%%%%%%%%%%%%%%%%
\begin{abstract}
Multiple scattering theory is applied to the study of clusters of point-like scatterers attached to a thin elastic plate and arranged in quasi-periodic distributions. Two type of structures are specifically considered: the twisted bilayer and the quasi-periodic line. The former consists in a couple of two-dimensional lattices rotated a relative angle, so that the cluster forms a moir\'e pattern. The latter can be seen as a periodic one-dimensional lattice where an incommensurate modulation is superimposed. Multiple scattering theory allows for the fast an efficient calculation of the resonant modes of these structures as well as for their quality factor, which is thoroughly analyzed in this work. The results show that quasi-periodic structures present a large density of states with high quality factors, being therefore a promising way for the design of high quality wave-localization devices.
\end{abstract}
%%%%%%%%%%%%%%%%%%%%%%%%%%%%%%%%%%%%%%%%%%%%%%%%%%%%%%%%%%%%%%%%%%%%%%%%%%%%%%%%%%%%%%%%%%%
\maketitle
%%%%%%%%%%%%%%%%%%%%%%%%%%%%%%%%%%%%%%%%%%%%%%%%%%%%%%%%%%%%%%%%%%%%%%%%%%%%%%%%%%%%%%%%%%%
\section{Introduction}
The study of quasi-periodic structures has a long story in all domains of physics\cite{PhysRevLett.53.1951,bindi2015natural,PhysRevLett.110.176403,man2005experimental,PhysRevB.94.205437,PhysRevLett.121.126401,hays1976variations,della2005band}, although recently an increasing interest has emerged due to the extraordinary properties of twisted bilayer graphene \cite{cao2018unconventional,polshyn2019large,moon2013optical}.

In the realm of classical waves (mainly photonic and acoustics), different quasi-periodic structures have been recently studied in one and two dimensions, and localized and robust interface states have been found both theoretically and experimentally \cite{rosa2021topological,beli2021mechanics,kuznetsova2021localized,vardeny2013optics,PhysRevX.6.011016,PhysRevLett.122.095501,ni2019observation,doi:10.1063/5.0042294,pal2019topological}. The theoretical analysis of quasi-periodic distributions of scatterers is extremely challenging, since these contain a large number of scatterers and some periodicity might only be retrieved in a higher-dimensional space, where the analysis might be simplified making use of a single unit cell \cite{Duneau+Katz1985}. Recently, we found that multiple scattering theory is a reliable method for the study of quasi-periodic arrangement of scatterers \cite{marti2021dipolar,marti2021edge}, and we showed that these structures present a large density of states.

However, the analysis of these states is incomplete if the imaginary part of the frequency is not taken into account, since open systems have finite mean-life resonances whose quality can be more than relevant for their use as wave-trapping devices. Consequently, a deeper analysis of these structures has yet to be done.

In this work we will revisit the structures studied in [\onlinecite{marti2021dipolar,marti2021edge}] to further analyze the resonances found therein. Multiple scattering theory allows for the analysis of the resonances' quality in finite clusters of scatterers. This can be done mainly by two methods, based on the analysis of the response of the system to real or complex frequencies. Both methods are similar for high quality resonances but only the latter is accurate for low quality factor resonances, as will be shown in the following sections.

The letter is organized as follows. After this introduction, in section \ref{sec:mst} we will present a brief description of the use of multiple scattering theory for the analysis of resonant frequencies of finite clusters of scatterers. Then, section \ref{sec:1D} will analyze the quality factor of the eigenmodes of the quasi-perioidc line of scatterers, and in section \ref{sec:TB} the same analysis will be performed to the two-dimensional twisted bilayer structure. Finally, section \ref{sec:summary} will summarize the work.

\begin{figure}[h!]
\centering \epsfig{file=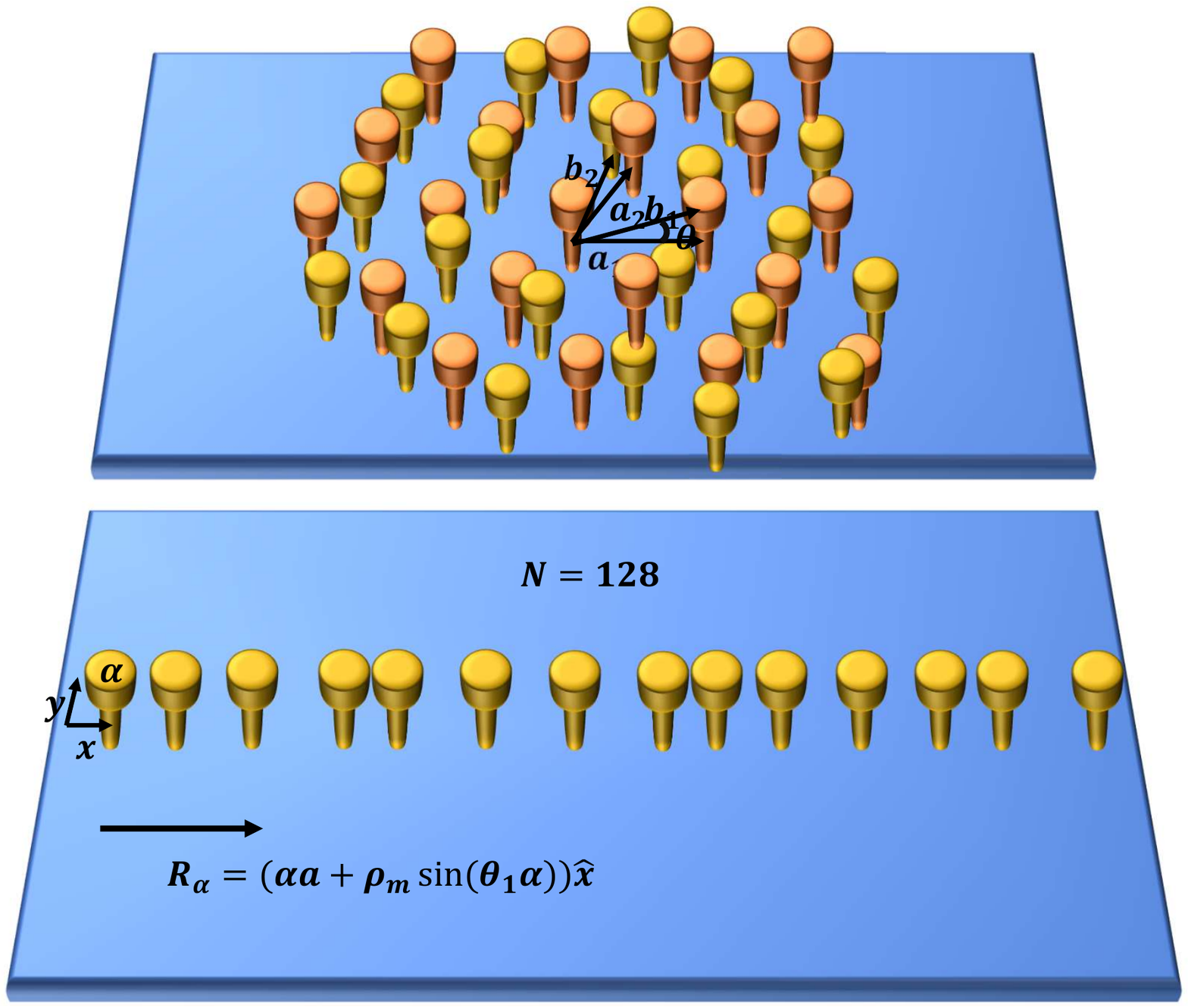, width=0.5\textwidth}
\caption{\label{Figura1} Schematics of the two structures analyzed in this work: A twisted bilayer configuration (upper panel) and an aperiodic line of scatterers (lower panel). The scatterers consist of spring-mass resonators attached to a thin elastic plate by means of a point-like contact. The geometrical parameters are indicated in the drawing (see text for further explanation).} 
\end{figure}
%%%%%%%%%%%%%%%%%%%%%%%%%%%%%%%%%%%%%%%%%%%%%%%%%%%%%%%%%%%%%%%%%%
\section{Resonant modes by multiple scattering theory}
\label{sec:mst}
Given a cluster of $N$ point scatterers attached to a thin elastic plate at positions $\bm{R}_\alpha$, for $\alpha=1\ldots N$, the eigenmodes of such a cluster are defined as the zeros of the determinant of the multiple scattering matrix $M$ when no incident field is present\cite{marti2021dipolar,marti2021edge}. This matrix is defined as
\beq
	M_{\alpha\beta} = \delta_{\alpha\beta}t_{\alpha}^{-1} - G(\mathbf{R}_{\alpha} - \mathbf{R}_{\beta}),
\label{EqDefM}
\eeq
where the Green's function $G(\fr)$ for flexural waves is given by
\beq
	G(\fr) = \frac{i}{8k^2}[H_0(kr)-H_0(ikr)],
\eeq
and the scalar quantity $t_\alpha$ defines the impedance of each scatterer \cite{marti2021dipolar,marti2021edge,torrent2013elastic, lera2019valley,karlos2022nonlinear}.

The matrix $M$ is a function of frequency via both $t_\alpha$ and $G(\fr)$. However the determinant is never zero for a real frequency for an open system (spectra of open systems depart from the real line and populate the complex plane \cite{kim2009computation}), as it is the case of our finite cluster. Therefore, in order to find the eigenmodes of the cluster we should solve the equation
\beq
\det M(\omega)=0
\eeq
for a complex frequency $\omega$. The real part of such a complex frequency corresponds to the peak expected upon excitation of the cluster while the imaginary part is related with the quality factor of the eigenmode, i.e., the larger the imaginary part, the lower the quality factor. 

Finding the complex zeros of a determinant is in general a difficult problem, since the determinant itself is complex-valued. However, an approximate value for the real part of the eigenfrequency can be obtained as those frequencies which minimize the value of the determinant, or, more efficiently, those frequencies which minimize the smallest eigenvalue of the matrix $M$. If the quality of the resonance is high it will have a small imaginary part, so that the complex zero can be found around the approximate real frequency which minimizes the smallest eigenvalue of $M$. Obviously, the closer the smallest eigenvalue to zero, the smaller the imaginary part of the eigenfrequency.

If the analysis is performed only analyzing the behaviour of the minimum eigenvalue as a function of the real part of the frequency, the quality factor can be found from the response in the neighborhood of the minimum. Thus, the quality factor of a resonance is defined as
\beq
    Q = \frac{f_0}{\delta f},
\eeq
with $f_0$ being the resonant frequency and $\delta f$ the full width half maximum. 

Both methods will be applied below for the characterization of the eigenmodes of the clusters.

\section{One-dimensional array of scatterers}  
\label{sec:1D}

We will consider first the analysis of eigenfrequencies of one-dimensional arrays of scatterers. Let us assume that a cluster of $N$ scatterers are arranged in positions $\mathbf{R}_{\alpha}$ as shown in figure \ref{Figura1}, lower panel, such that
\beq
    \mathbf{R}_\alpha = a\alpha + \rho_m sin(\alpha\theta),
\eeq
with $a$ being the lattice constant, $\rho_m$ the radius of the modulation circle and $\theta$ the angle rotated in the circle (as defined in [\onlinecite{apigo2019observation}]). This type of arrangement allows to define periodic or aperiodic clusters depending on the nature of the $\theta$ parameter. 

\begin{figure}[h!]
\centering \epsfig{file=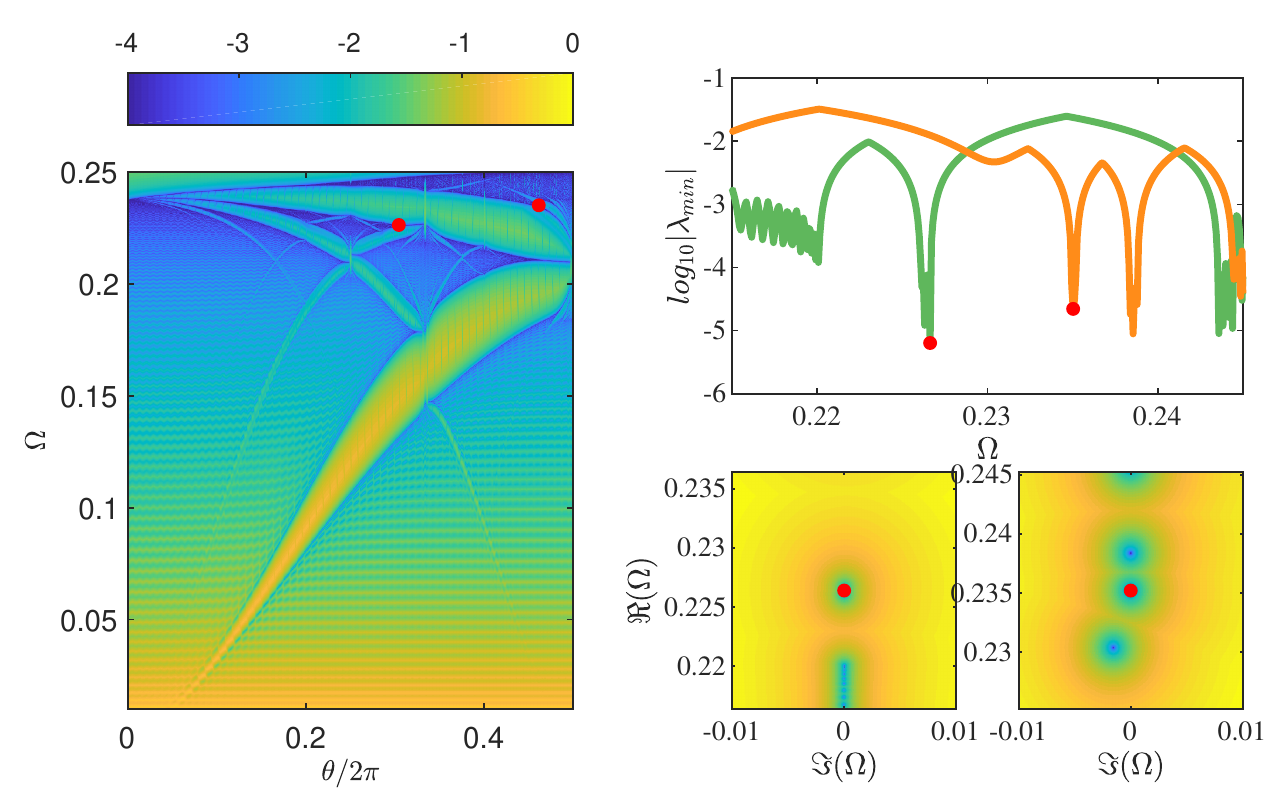, width=0.5\textwidth}
\caption{\label{Figura5} Left panel: evolution of the minimum eigenvalue of the multiple scattering matrix ($M$) as a function of the modulation parameter of an one-dimensional array of scatterers and the normalized frequency of the system (real frequency). The scatterers' properties are $\gamma_\alpha = 100$ and $\Omega_{\alpha}·a/\pi = 0.25$. Upper right panels: evolution of the minimum eigenvalue as a function of the normalized frequency for two spatial configurations, corresponding to $\theta = 0.26\degree$ and $\theta = 0.45\degree$. Lower right panels: these two maps show the evolution of the minimum eigenvalue as a function of the normalized frequency of the system in the complex plane.} 
\end{figure}

Figure \ref{Figura5}, left panel shows the evolution of the minimum eigenvalue of the $M$ matrix as a function of the normalized frequency of the system (only real component) and the modulation parameter $\theta$. The diagram shows the well known Hofstadter's butterfly \cite{PhysRevB.14.2239}, and it was previously found for these systems\cite{marti2021edge}. This structure is characterized by a set of gaps without modes all over the spectrum, defining the contour of the butterfly (yellow and green regions in figure \ref{Figura5} left panel). The upper right panel of figure \ref{Figura5} shows the frequency evolution of the smallest eigenvalue for $\theta = 0.26\degree$ (green line) and $\theta = 0.45\degree$ (orange line). Two points, highlighted in red, belong to two different zones where two bandgaps are approaching each other. These bandgaps will collapse as the modulation parameter increases; however, before this occurs, they confine a state in between. The green line shows a resonance in the middle of two bandgaps, and at low frequencies several resonances appear. The complex frequency map shown below also depicts these minima: the imaginary part that must be added in order to obtain the cancellation of the eigenvalue is negligible. Concerning the second configuration, one minimum has been chosen (red point). However, the chosen window lets us appreciate four different minima in the eigenvalue; three of them having high quality factor, while another one having low quality factor (near $\Omega = 0.23$). The complex frequency map below also shows these four minima: the three in the upper part correspond to the high quality resonances; they are at the center of the map, with almost null imaginary frequency component. Meanwhile, the fourth resonance is shifted at left, stating the need of a higher imaginary component for achieving the cancellation of the determinant. 

These modes tend to localize at the edge of the structure, so that by adding the mirror symmetric cluster we will increase the robustness of the mode, as will be shown below. They appear as a consequence of having a finite structure. If we were able to simulate the infinite structure, we would not be able to find these modes inside the gap. This result can be approached by only evaluating our structure for modulation parameters such that:

\begin{equation}
    \theta = \frac{n}{N}, n = 1,...,N,
\end{equation}

with $N$ being the number of scatterers in the finite cluster, as explained in \onlinecite{gupta2020dynamics}. Edge states became interface states by the addition of mirror symmetry. 

\begin{figure}[h!]
\centering \epsfig{file=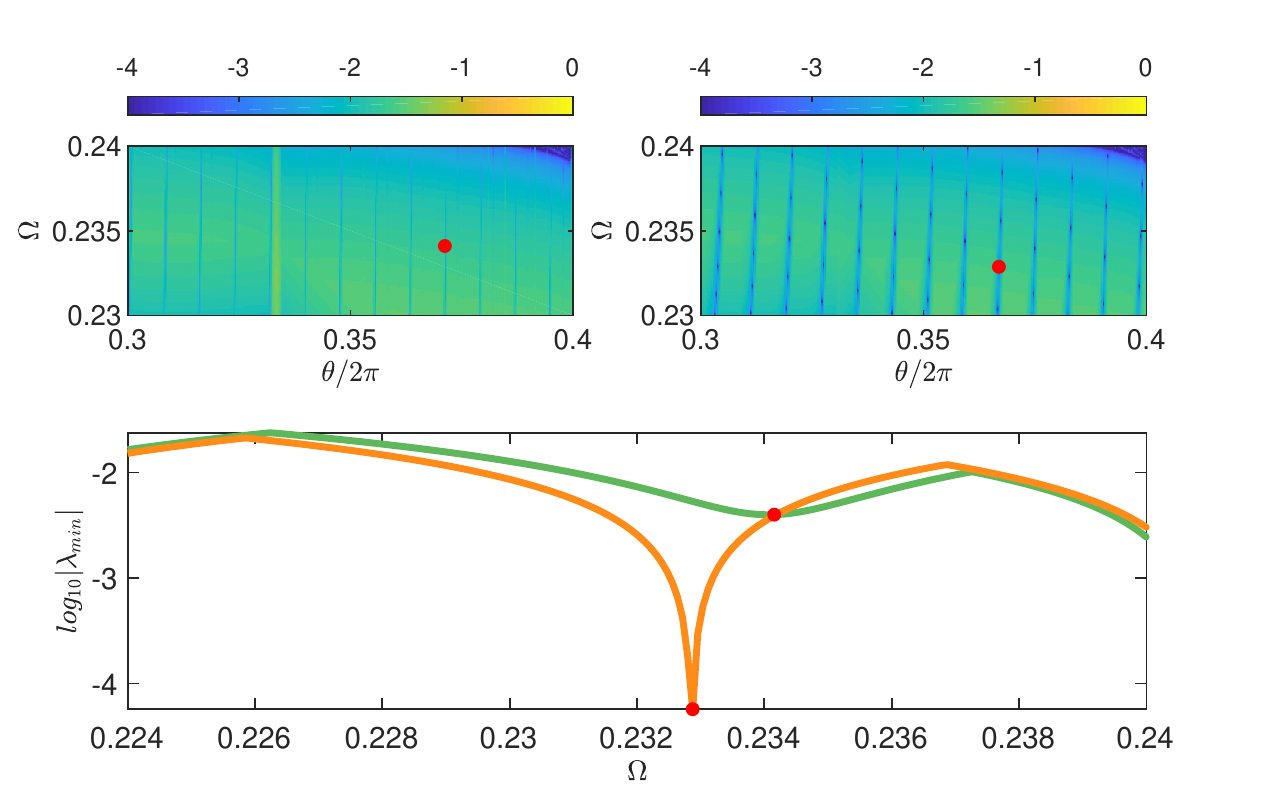, width=0.5\textwidth}
\caption{\label{Figura6} Resonance comparison between modes found in the gaps of the Hofstadter's butterfly for a one-dimensional array of scatterers  (upper left panel) and bandgap modes found for a one-dimensional array of scatterers with mirror symmetry (upper right pannel). Lower panel: evolution of the minimum eigenvalue as a function of the normalized frequency for both structures with a given spatial configuration. As can be seen, the mirror-symmetric structure presents a sharper resonance compared with the simpler structure.} 
\end{figure}

Figure \ref{Figura6}, upper panels, show the same insight of the Hofstadter butterfly but the left map corresponds to the linear array of scatterers without mirror symmetry, while the second one corresponds to the addition of mirror symmetry to the structure. These insights are centered in a bandgap; therefore, the vertical lines that appear in both maps (in the left one they are hardly visible, while they are clearer in the right one) correspond to modes inside the gap. One point has been chosen from each map (they have not exactly the same properties, due to the fact that vertical lines are slightly shifted when adding the mirror symmetry to the structure). Figure \ref{Figura6} lower panel shows the evolution of the eigenvalue as a function of the normalized frequency for the spatial configuration of the red points. Whereas the simple structure shows a low quality factor resonance with a minimum far from zero, the mirror-symmetric structure presents a narrower resonance (high quality factor) and the minimum is two orders of magnitude smaller than the former one.  

\begin{figure}[h!]
\centering \epsfig{file=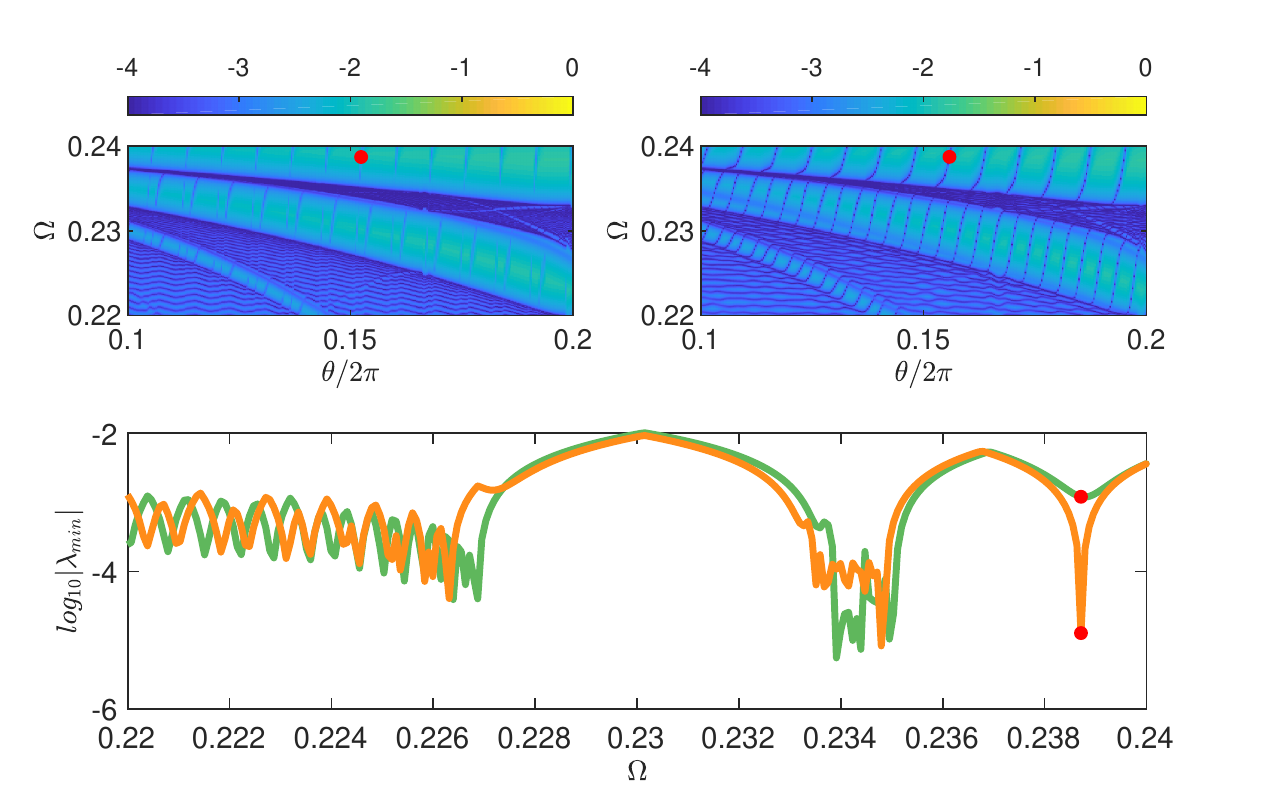, width=0.5\textwidth}
\caption{\label{Figura7} Another example of comparison between one-dimensional arrays of scatterers (upper left panel) and one-dimensional arrays of scatterers with mirror symmetry (upper right panel). Again, the lower panel shows the evolution of the minimum eigenvalue for a given spatial configuration, showing that the resonance near $\Omega = 0.24$ is sharper in the case of the mirror-symmetric structure.} 
\end{figure}

Figure \ref{Figura7} shows another example of the improvement in the bandgap modes due to the addition of symmetry. In this case, the chosen spatial configuration is found inside one of the smaller bandgaps of the Hofstadter's butterfly. However, the result is similar: the evolution of the eigenvalue as a function of the frequency shows that the resonance inside the bandgap gets narrower when adding spatial symmetry to the system.

As discussed before, we expect a small imaginary part for these two resonances when the zeros are searched in the complex plane, since the eigenvalue is very close to zero.
%%%%%%%%%%%%%%%%%%%%%%%%%%%%%%%%%%%%%%%%%%%%%%%%%%%%%%%%%%%%%%%%%%
\section{Twisted bilayers}
\label{sec:TB}
In this section we will perform a similar analysis but for the twisted bilayer analyzed in [\onlinecite{marti2021dipolar}]. This analysis is motivated by the fractal energy landscapes of twisted bilayers\cite{dean2013hofstadter}, similarly to the Hofstadter butterfly spectrum of the finite cluster studied previously. 
 We will build clusters of scatterers by the superposition of two identical periodic lattices with a relative angle between their lattice vectors. Thus, defining the first lattice vectors as $\mathbf{a_l} = a_{lx} + ia_{ly}$, for $l = 1,2$, and similarly for the second lattice ($\mathbf{b_l} = b_{lx} + ib_{ly}$), they will be simply related as $\mathbf{b_l} = \mathbf{a_l} e^{i\theta_0}$, being $\theta_0$ the twisting angle between both lattices. Scatterers in the cluster are located at positions $R_{a} = n_1\mathbf{a_1} + n_2\mathbf{a_2}$ and $R_{b} = m_1\mathbf{b_1} + m_2\mathbf{b_2}$, being $n_i,m_i$ integer numbers.

\begin{figure}[h!]
\centering \epsfig{file=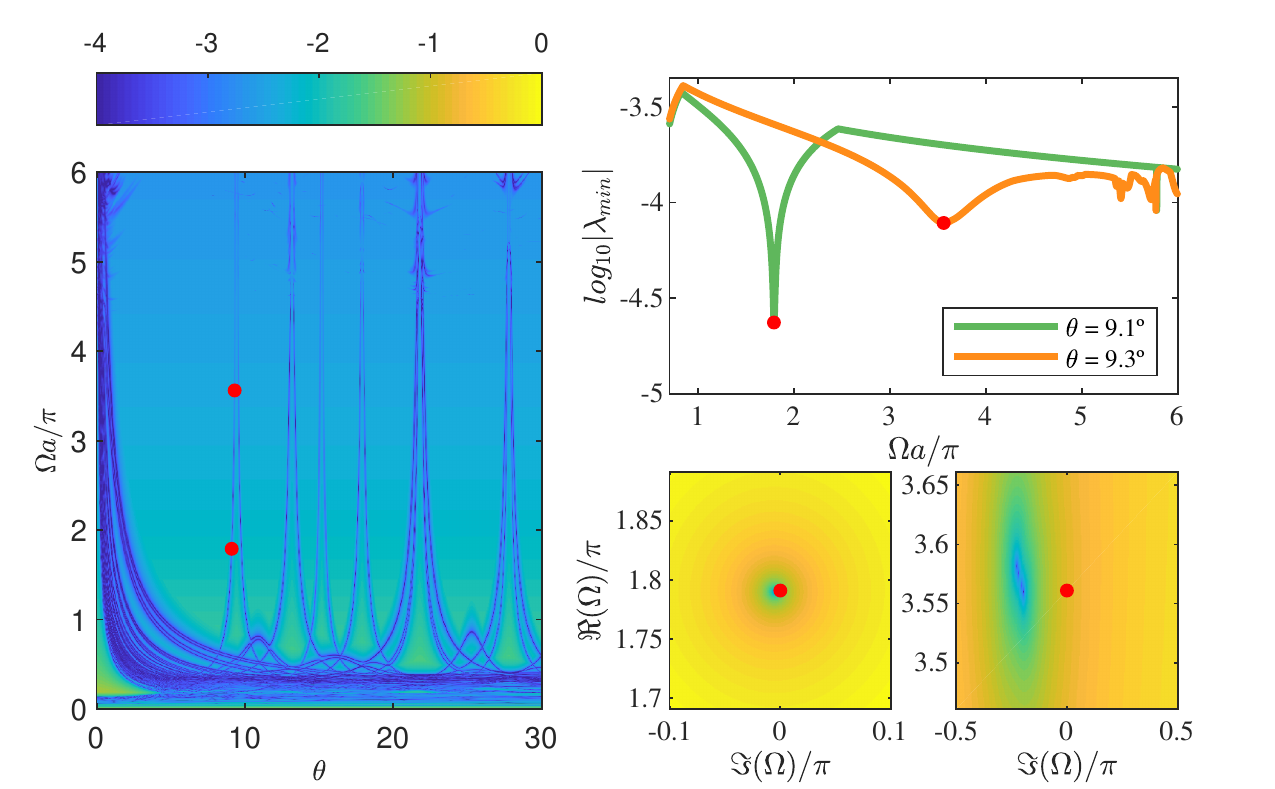, width=0.5\textwidth}
\caption{\label{Figura2}{ Left panel: evolution of the minimum eigenvalue of the multiple scattering matrix ($M$) as a function of the twisting angle of the twisted bilayer structure and the normalized frequency of the system (real frequency). The scatterers' properties are $\gamma_\alpha = 200$ and $\Omega_{\alpha}·a/\pi = 20$. The original lattices forming the twisted bilayer have a triangular arrangement. Upper right panel: evolution of the minimum eigenvalue as a function of the normalized frequency for two spatial configurations, corresponding to $\theta = 9.1\degree$ and $\theta = 9.3\degree$. Lower right panels: these two maps show the evolution of the minimum eigenvalue as a function of the normalized frequency of the system in the complex plane.}} 
\end{figure}

Figure \ref{Figura2}, left panel, shows the evolution of the minimum eigenvalue of the $M$ matrix as a function of the normalized frequency of the system and the twisting angle of the bilayer structure. Blue lines in the map correspond to configurations where the eigenvalue approaches zero. The individual lattices forming the twisted bilayer have triangular arrangement, and the properties of the scatterers are constant ($\Omega_{\alpha} a / \pi = 20$ and $\gamma_{\alpha}=200$). Triangular lattices have $\pi/6$ symmetry; therefore, only the twisting angles between $0\degree$ and $30\degree$ have been shown. As explained in [\onlinecite{marti2021dipolar}], modes are located around commensurate angles of the twisted bilayer and also near small twisting angles. Two points have been chosen in the map: they have been highlighted in red. In the upper right panel, the evolution of the minimum eigenvalue with frequency is shown for the two spatial configurations of the highlighted points: they correspond to $\theta = 9.1\degree$ and $\theta = 9.3\degree$. Both curves present a resonance, being the red point the minimum of the resonance. However, while the first curve shows a sharp resonance with a high quality factor ($Q = 81.4$), the second curve has a broader resonance with low quality factor ($Q = 4.6$).

As explained before, once the resonant frequency has been found, a detailed analysis in the complex frequency plane can be done. Figure \ref{Figura2} lower panels show the evolution of the minimum eigenvalue as a function of the complex normalized frequency of the system. Both maps are centered at the resonant frequencies of the chosen modes (red points). The first case, which presented a sharp resonance, shows the minimum in the center of the complex map, with a slight shift with respect to the red point. This means that a small imaginary frequency term is needed in order to achieve the cancellation of the minimum eigenvalue and therefore the cancellation of the determinant of the $M$ matrix. As for the second mode, its complex map shows a minimum shifted to the left part of the map. It can be seen that the minimum is no longer a point but a distributed structure, being this an artifact produced by the discretization of the parameters in the simulation so that an artificial breaking of the degeneracy is found. 

\begin{figure}[h!]
\centering \epsfig{file=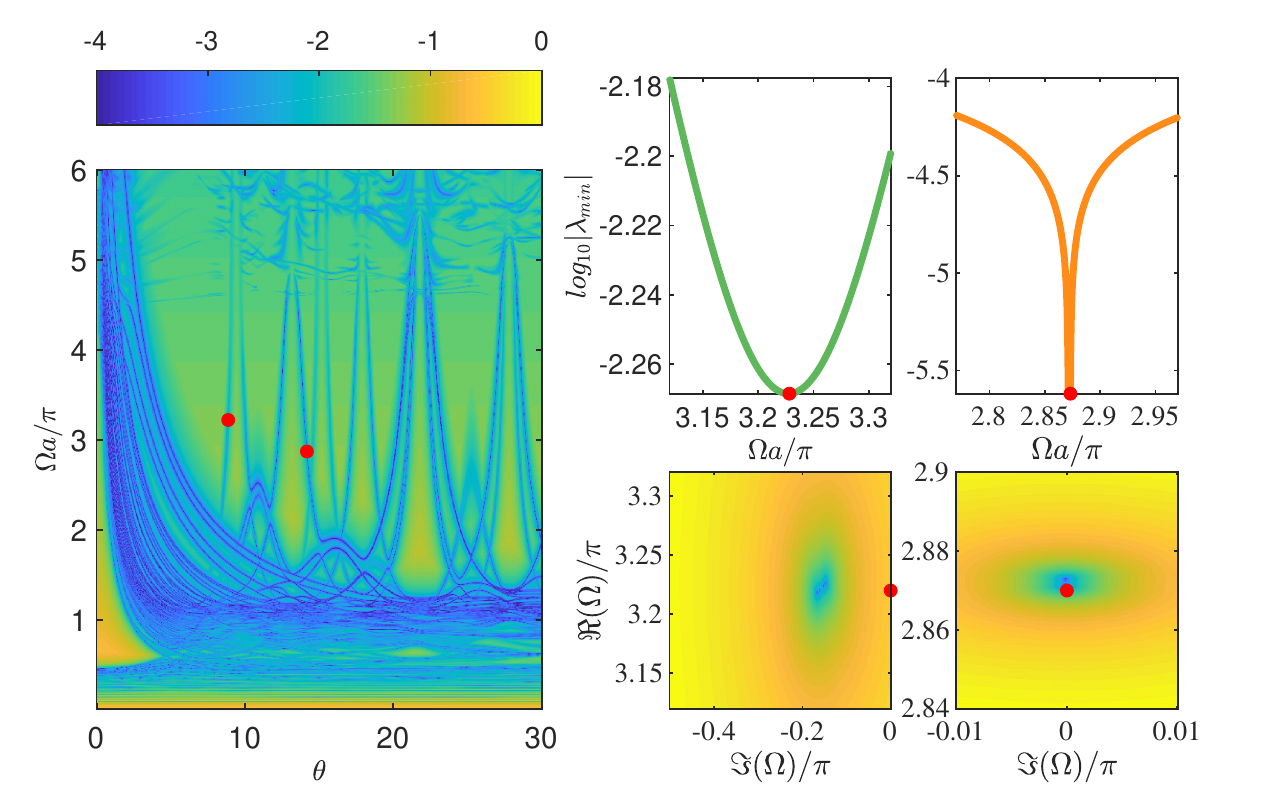, width=0.5\textwidth}
\caption{\label{Figura3}{ Left panel: evolution of the minimum eigenvalue of the multiple scattering matrix ($M$) as a function of the twisting angle of the twisted bilayer structure and the normalized frequency of the system (real frequency). The scatterers' properties are $\gamma_\alpha = 20$ and $\Omega_{\alpha}·a/\pi = 20$. The original lattices forming the twisted bilayer have a triangular arrangement. Upper right panel: evolution of the minimum eigenvalue as a function of the normalized frequency for two spatial configurations, corresponding to $\theta = 8.86\degree$ and $\theta = 14.17\degree$. Lower right panels: these two maps show the evolution of the minimum eigenvalue as a function of the normalized frequency of the system in the complex plane.}} 
\end{figure}

By changing the normalized mass of the scatterers, we are able to change their impedance, and then change the behavior of the cluster. Figure \ref{Figura3} left panel shows the evolution of the minimum eigenvalue of the cluster as a function of the twisting angle and the real frequency of the system. This is the same map as in figure \ref{Figura2}, but this time the normalized mass is $\gamma_{\alpha} = 20$. What can be seen from this map is that by reducing the impedance of the scatterers, the behavior at low frequency becomes more complex. Compared to the case $\gamma_{\alpha} = 200$, this one shows more modes for small angles. Furthermore, dipolar behavior near commensurate angles, which was clear in the precedent case whenever $\Omega a /\pi > 1$, is now more confusing. Only when $\Omega a / \pi > 2.5$ dipolar modes appear. Again, two configurations in this map have been chosen, highlighted with red points in the map. Their resonances as a function of the frequency are shown in the upper right panels of figure \ref{Figura3}. In this case, each resonance is shown in a different graph. We have decided to split it due to the complexity of the curve with frequency. Therefore, the graphs shown are centered in the resonance, not as in figure \ref{Figura2}, where all the frequency range was plotted. The first resonance, which corresponds to a spatial configuration of $\theta = 8.86\degree$ and a resonant frequency of $\Omega a /\pi = 3.23$, shows a broad resonance behavior, having a low quality factor ($Q = 5.98$). This result is in accordance with what it is seen in the complex map below. The minimum in the complex map appears shifted with respect to the red point. The imaginary part needed is higher than in the precedent cases. The second one ($\theta = 14.17\degree$ and $\Omega a / \pi = 2.88$) shows a sharper behavior, and thus the complex map shows that the minimum is almost at the center of the map, near the brown point. These two cases have similar behavior to those shown for the $\gamma_{\alpha} = 200$ twisted bilayers. 

\begin{figure}[h!]
\centering \epsfig{file=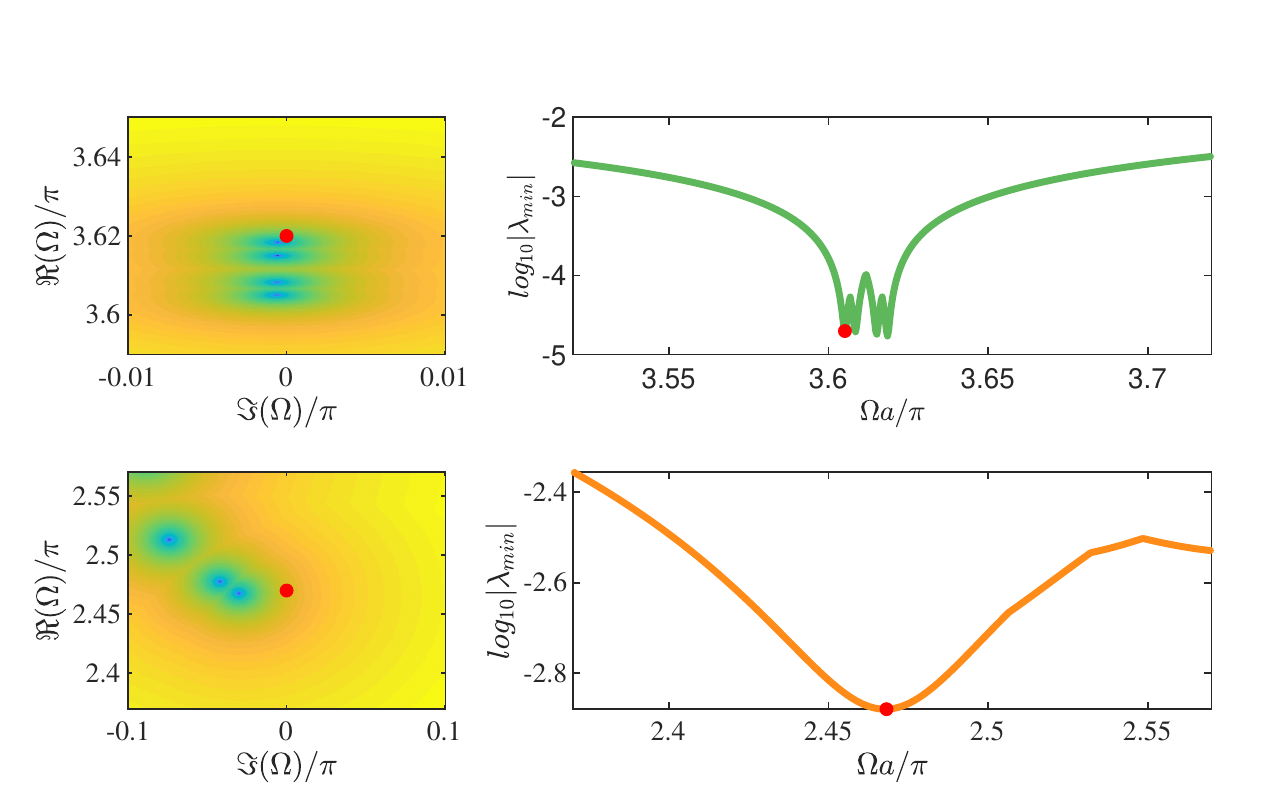, width=0.5\textwidth}
\caption{\label{Figura4} Two more resonances on the $\gamma_{\alpha}=20$ twisted bilayers. The first one ($\theta = 12.47\degree$), shown above, has been split in four sharp peaks. The second resonance ($\theta = 0.65\degree$), shown below, is a wide one. However, the complex frequency map shows the presence of different minima in the region where the resonance is found.} 
\end{figure}

Finally, other resonances with more complicated behavior can be found in this second structure (twisted bilayers with $\gamma_{\alpha} = 20$). Figure \ref{Figura4} shows two more resonances (evolution of the minimum eigenvalue with real frequency and the complex frequency map around the resonance point). The one shown in the upper panels has $\theta = 12.47\degree$ and $\Omega a / \pi = 3.62$, which is located in the dipolar modes of the left panel in figure \ref{Figura3}. In this case, the resonance is split in four sharp minima. The same behavior is seen in both the real frequency line and the complex map. In the latter, four minima appear with the same imaginary component, which means that the quality factor of these resonances is the same. As for the second mode, the spatial configuration is $\theta = 0.65\degree$, and the resonant frequency is $\Omega a /\pi = 2.47$. This point belongs to the left part of figure's \ref{Figura3} left panel, in other words, to the small angle condition, where all the scatterers in the cluster are close to each other. The behavior of the eigenvalue with real frequency shows a single minimum with a wide resonance and a low quality factor ($Q = 20.58$). However, the insight in the complex plane reveals a more complex behavior, with four different minima with different imaginary components and even different real frequencies. As we have found in the aperiodic line of scatterers, this structure also presents some regions with a high density of states.

% DISCUSION SOBRE LOS DATOS NUMERICOS DE LAS "FLECHAS" EN FRECUENCIA

\section{Summary}
\label{sec:summary}
In summary, we have analyzed the eigenfrequencies of finite clusters of scatters arranged in quasi-periodic distributions by multiple scattering theory. Unlike previous works, in which only the position of these modes has been considered, we have focused our study in the quality of these modes, which is a more relevant parameter given the great amount of resonances that these structures present. Two types of clusters have been studied, linear distributions in which the modulation of the distance between scatterers creates the aperiodicity and twisted bilayers, in which a moir\'e pattern is formed. The former type of clusters presents localized modes at its edges, although the quality of these modes is remarkably enhanced at the interface beween the cluster and its mirror symmetric structure. The latter type of clusters shows as well high quality modes, although a strong variation of this quality can be found through the configuration map, which shows that the analysis of the quality of these modes is more than relevant for their use in operating devices.
%%%%%%%%%%%%%%%%%%%%%%%%%%%%%%%%%%%%%%%%%%%%%%%%%%%%%%%%%%%%%%%%

\section*{Acknowledgments}
D.T. acknowledges financial support through the ``Ram\'on y Cajal'' fellowship, under Grant No. RYC-2016-21188, and from the Ministry of Science, Innovation, and Universities, through Project No. RTI2018- 093921-AC42. M. M.-S. acknowledges financial support through the FPU program, under Grant No. FPU18/02725. Research supported by DYNAMO project (101046489), funded by the European Union. Views and opinions expressed are however those of the authors only and do not necessarily reflect those of the European Union or European Innovation Council. Neither the European Union nor the granting authority can be held responsible for them.
\section*{Code availability}
Numerical calculations are done with home-made scripts, which are available upon request to the authors.
%%%%%%%%%%%%%%%%%%%%%%%%%%%%%%%%%%%%%%%%%%%%%%%%%%%%%%%%%%%%%%%%

\appendix

%\bibliography{bibliography}

\end{document}